\documentclass[onecolumn, 12pt]{IEEEtran}
\usepackage{graphicx}
\usepackage{cite}
\usepackage{amsmath}
\usepackage{amsthm}
\usepackage{mathtools}
\usepackage{amssymb}
\usepackage{subfigure}
\usepackage{setspace}
\usepackage{bbm}
\usepackage[justification=centering]{caption}

\newcommand{\al}{\left[\frac{\lambda (1-\mu_1)}{(1-\lambda) \mu_1} \right]^M}

\IEEEoverridecommandlockouts

\ifCLASSINFOpdf
\else
\fi

\begin{document}

\title{Throughput of a Cognitive Radio Network under Congestion Constraints: A Network-Level Study}
\author{\IEEEauthorblockN{Nikolaos Pappas\textsuperscript{*}, Marios Kountouris\textsuperscript{\dag}}\\
\IEEEauthorblockA{\textsuperscript{*} Department of Science and Technology, Link\"{o}ping University,\\ Norrk\"{o}ping SE-60174, Sweden \\
\textsuperscript{\dag} Sup\'{e}lec, Department of Telecommunications, Gif-sur-Yvette, France\\
Email: nikolaos.pappas@liu.se, marios.kountouris@supelec.fr
}

\thanks{This work has been partially supported by the People Programme (Marie Curie Actions) of the European Union's Seventh Framework Programme FP7/2007-2013/ under REA grant agreement no.[612361] -- SOrBet and by the ERC Starting Grant 305123 MORE (Advanced Mathematical Tools for Complex Network Engineering).}

}
\maketitle

\begin{abstract}
In this paper we analyze a cognitive radio network with one primary and one secondary transmitter, in which the primary transmitter has bursty arrivals while the secondary node is assumed to be saturated (i.e. always has a packet waiting to be transmitted). The secondary node transmits in a cognitive way such that it does not impede the performance of the primary node. We assume that the receivers have multipacket reception (MPR) capabilities and that the secondary node can take advantage of the MPR capability by transmitting simultaneously with the primary under certain conditions. 
We obtain analytical expressions for the stationary distribution of the primary node queue and we also provide conditions for its stability. Finally, we provide expressions for the aggregate throughput of the network as well as for the throughput at the secondary node.
\end{abstract}

\section{Introduction}
Cognitive radio communication provides an efficient means of sharing radio spectrum between users having different priorities \cite{zhao:survey}. 
The term ``Cognitive Radio" was first introduced by Mitola in the 1990s to take advantage the highly under-utilized scarce wireless spectrum \cite{b:Mitola}.
The high-priority transmitter, usually termed as primary, is allowed to access the channel whenever it is needed, while the low-priority node, coined as secondary, is required to make a decision on its transmission and access the channel opportunistically based on the spectrum occupancy and the primary user transmission.

In this paper, we consider a cognitive radio network consisting of a primary user with bursty arrivals and a secondary user with saturated traffic, as shown in Fig.~\ref{fig:model}.
We obtain the aggregate throughput of the network for a cognitive access protocol on the general multipacket reception (MPR) channel model. The MPR channel captures the effects of fading, attenuation, and interference at the physical layer in a more efficient way than the traditional collision channel model, as in the former a transmission may succeed even in the presence of interference \cite{ghez:stability, tong:multipacket, naware:stability, b:PappasMPR}.

The secondary transmitter can take advantage of the MPR capability by transmitting simultaneously with the primary node under certain conditions. We slightly modify the cognitive access protocol proposed in \cite{rong:cooperation, kompella:stable, b:Pappas-JCN}, in which the secondary node not only utilizes the idle periods of the primary node, but also competes with the primary node by randomly accessing the channel when the queue size of the primary node is below a congestion limit. If the primary node queue size exceeds that limit, then the secondary node is not allowed to transmit. The congestion limit is a way to ensure that the secondary node will not harm the primary node more than a certain level.

To position our contribution with respect to the literature, we provide below a brief background review. In~\cite{b:Sadek}, a novel cognitive multiple-access strategy in the presence of a cooperating relay is proposed. That work was among the first that introduced the notion of network-level cooperation, i.e. cooperation without any physical layer processing. In~\cite{b:Pappas-ISIT}, the notion of partial (or probabilistic) network-level cooperation is introduced, where probabilistic cooperation means that under certain conditions in the network, the relay may accept a packet from the source.  
In~\cite{b:Krikidis-CognitiveMA}, the authors study the impact, from a network-layer perspective, of having a single cognitive radio transmitter- receiver pair sharing the spectrum with multiple primary users wishing to communicate to a single receiver in a multi-access channel (MAC). In~\cite{b:Pappas-JCN}, an opportunistic multiple access protocol is proposed, in which the priorities among the users are observed in order to better utilize the limited energy resources. Owing to the MPR capability, the secondary node not only utilizes the idle slots, but can also take advantage of the additional reception by transmitting along with the primary node in a random access way that does not adversely affect the quality of the communication over the primary link.

In this paper, we first analyze the queue characteristics of the primary transmitter. Specifically, we model the queue as a discrete time Markov Chain and we obtain its stationary distribution. Furthermore, we provide the conditions for the stability of the queue and we characterize the throughput experienced by the secondary transmitter, as well as the aggregate throughput of the network.

The rest of the paper is organized as follows: in Section~\ref{sec:model}, we describe the system model including the network model and the cognitive protocol. In Section~\ref{sec:Analysis}, we include the analysis for the primary node queue, and numerical results are presented in Section~\ref{sec:Results}. Finally, Section~\ref{sec:conclusions} concludes our work.

\section{System Model} \label{sec:model}
\subsection{Network Model}
\begin{figure}[h]
\centering
\includegraphics[scale=1.3]{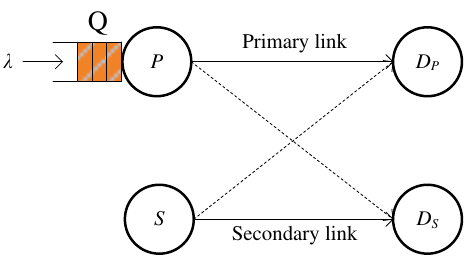}
\caption{A cognitive network with a primary and a secondary transmitter. The primary node has bursty traffic whereas the secondary has a saturated queue.}
\label{fig:model}
\end{figure}

We consider a cognitive radio network of two source-destination pairs, as shown in Fig.\ref{fig:model}. The time is slotted and each packet transmission occupies one time slot.
The primary source has an infinite capacity queue $Q$ for storing the arriving packets of fixed length. The arrival process at the primary transmitter is modeled as a Bernoulli process with average rate $\lambda$ packet per slot. The secondary node queue is assumed to be saturated (always backlogged), i.e. it always has a packet waiting to be transmitted.

\subsection{Physical Layer Model}
The MPR channel model is a generalized form of the packet erasure model. At the receiver side, a packet can be decoded correctly by the receiver if the received SINR exceeds a certain threshold. More precisely, suppose that we are given a set $T$ of nodes transmitting during the same time slot. Let  $P_{rx}(i,j)$ be the signal power received from node $i$ (where $i = P, S$) at node $j$ (where $j = D_P, D_S$), and let ${\rm SINR}(i,j)$ be the SINR determined by node $j$, i.e.
\begin{equation*}
{\rm SINR}(i,j)=\frac{P_{rx}(i,j)}{\eta_{j}+\sum_{k\in T\backslash\left\{i\right\}} {P_{rx}(k,j)}},
\end{equation*}
where $\eta_{j}$ denotes the receiver noise power at $j$. We assume that a packet transmitted by node $i$ is successfully received by $j$ if and only if ${\rm SINR}(i,j)\geq \gamma_{j}$, where $\gamma_{j}$ is a threshold characteristic of node $j$. Let $P_{tx}(i)$ be the transmit power at node $i$ and $r(i,j)$ be the distance between $i$ and $j$. The power received by node $j$ when node $i$ transmits is $P_{rx}(i,j)=A(i,j)g(i,j)$, where $A(i,j)$ is a random variable representing channel fading. We assume that the channel is subject to slow, flat fading, constant during a timeslot and independently varying from one timeslot to another. Under Rayleigh fading, it is known that $A(i,j)$ is exponentially distributed~\cite{b:Tse}. The received power factor $g(i,j)$ is given by $g(i,j)=P_{tx}(i)(r(i,j))^{-\alpha}$ where $\alpha$ is the pathloss exponent with $\alpha > 2$. The success probability of link $i-j$ when the transmitting nodes are in $T$ is given by
\begin{align*}
\label{eq:succprob}
P_{i/T}^{j}=\exp\left(-\frac{\gamma_{j}\eta_{j}}{v(i,j)g(i,j)}\right) \times \\
\times \prod_{k\in T\backslash \left\{i,j\right\}}{\left(1+\gamma_{j}\frac{v(k,j)g(k,j)}{v(i,j)g(i,j)}\right)}^{-1},
\end{align*}
where $v(i,j)$ is the parameter of the Rayleigh random variable for fading. 

For the sake of presentation, we denote $p_{1/1} = P_{P/P}^{D_P}$ and $p_{2/2} = P_{S/S}^{D_S}$ when only one transmitter is active. When both transmitters are active,
we have $p_{1/1,2} = P_{P/P,S}^{D_P}$ and $p_{2/1,2} = P_{S/P,S}^{D_S}$. Note that $p_{1/1} \geq p_{1/1,2}$ and $p_{2/2} \geq p_{2/1,2}$. 

\subsection{Cognitive Access Protocol} \label{sec:cogprot}
In this work, we build on the cognitive protocol proposed in~\cite{rong:cooperation} and we slightly modify it in the following way. The primary node transmits a packet whenever is backlogged. On the other hand, the secondary node transmits a packet by accessing the channel in a way that it does not deteriorate the performance of the primary node.
The secondary node monitors the status of the queue of the primary node, and when the queue is empty (thus the primary node is silent), the secondary node transmits with probability 1.
When the size of the queue in the primary node is between 1 and a threshold $M$,  the secondary node accesses the channel with probability $q$.

The service rate for the primary node in this case is denoted $\mu_1$ and is given by
\begin{equation} \label{eq:mu1}
\mu_1 = qp_{1/1,2} + (1-q)p_{1/1}.
\end{equation}

When the queue size is larger than $M$ then, the secondary node remains silent. In that case, the service rate of the primary node is $\mu_2$ and is given by
\begin{equation} \label{eq:mu2}
\mu_2 = p_{1/1}.
\end{equation}
The threshold $M$ plays the role of a congestion limit for the primary node, meaning that when the queue reaches this size then, the secondary node does not attempt to transmit any packet.

The throughput for the secondary user, $T_s$ is given by
\begin{equation} \label{eq:T_s}
T_{s}= \mathrm{Pr}\left(Q = 0 \right) p_{2/2} + \mathrm{Pr}\left(1 \leq Q \leq M \right)qp_{2/1,2}.
\end{equation}   

Thus, in order to compute $T_s$ we need to analyze the queue at the primary node. The analysis is given in the following section.

\section{Analysis of the Primary Node Queue} \label{sec:Analysis}
In order to characterize the performance in our system, we need to characterize the maximum stable throughput for the primary node and the average throughput for the secondary node. The maximum stable throughput is defined only for sources that are not backlogged, i.e. for sources with bursty arrivals, and is the rate measured in terms of packet/slot at which data is delivered from the transmitter to its intended receiver, while guaranteeing that the queue does not grow unbounded.

\begin{figure}[h]
\centering
\includegraphics[scale=0.35]{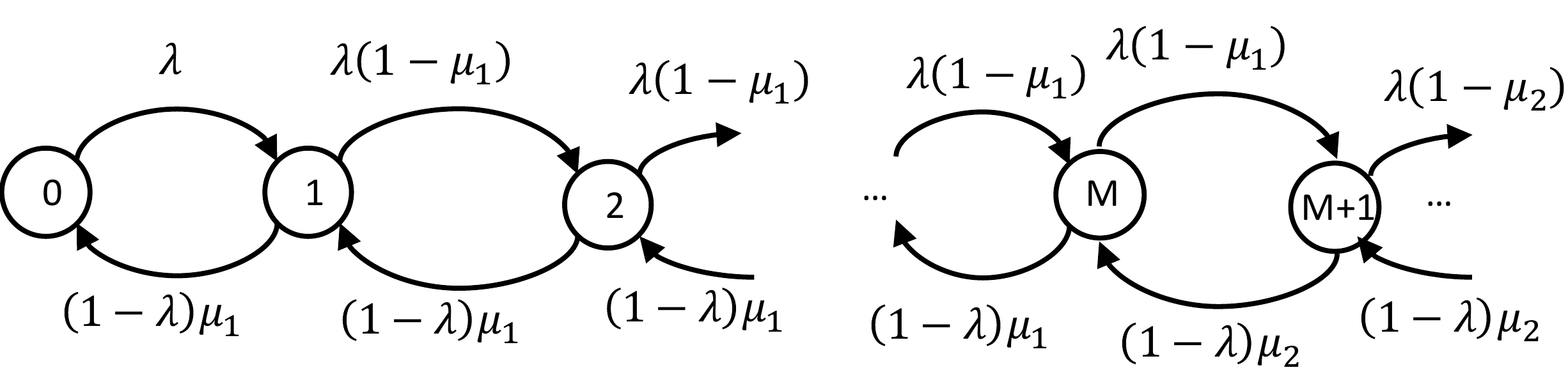}
\centering
\caption{The Discrete Time Markov Chain which models the queue evolution at the primary node.}
\label{fig:dtmc}
\end{figure}

The average throughput for the secondary node, given in (\ref{eq:T_s}), depends on the state of the primary node queue, hence we need to characterize $\mathrm{Pr}\left(Q=0 \right)$ and $\mathrm{Pr}\left( 1 \leq Q \leq M \right)$. We model the queue at the primary node as a discrete time Markov Chain (DTMC), which describes the queue evolution and is presented in Fig.~\ref{fig:dtmc}. Each state is denoted by an integer and represents the queue size at the primary node.

In order to compute the $\mathrm{Pr}\left(Q=0 \right)$ and $\mathrm{Pr}\left( 1 \leq Q \leq M \right)$, we utilize the balance equations
of the DTMC. The stationary distribution of the DTMC is denoted by $\pi$, where $\pi(i) = \mathrm{Pr}\left(Q=i \right)$ is the probability that the queue has $i$ packets when it is in steady state.

From the balance equations we obtain the following
\begin{equation*}
\lambda \pi(0) = (1-\lambda) \mu_1 \pi(1) \Leftrightarrow \pi(1) = \frac{\lambda}{(1-\lambda) \mu_1} \pi(0)
\end{equation*}

\begin{align*}
\left[\lambda (1-\mu_1) +(1-\lambda)\mu_1\right]\pi(1) = \lambda \pi(0)+(1-\lambda)\mu_1 \pi(2) \\
 \Leftrightarrow \pi(2) = \frac{\lambda^2 (1-\mu_1)}{(1-\lambda)^2 \mu_1^2} \pi(0)
\end{align*}

Summarizing, for $1 \leq i \leq M$ we have that
\begin{equation*}
\pi(i) = \frac{\lambda^i (1-\mu_1)^{i-1}}{(1-\lambda)^i \mu_1^i} \pi(0),
\end{equation*}

and for $i>M$ we obtain
\begin{equation*}
\pi(i) = \frac{\lambda^{M+i} (1-\mu_1)^M (1-\mu_2)^{i-1}}{(1-\lambda)^{M+i} \mu_1^{M} \mu_2^{i}} \pi(0).
\end{equation*}

The previous steady state probabilities are given as a function of $\pi(0)$, however it is known that
\begin{equation} \label{eq:sum}
\sum_{i=0}^{\infty} \pi(i)=1.
\end{equation}

From the previous expression, we derive the probability that the queue is empty and is given by
\begin{equation}
\mathrm{Pr}\left(Q=0 \right) = \frac{(\mu_1 - \lambda)(\mu_2 - \lambda)}{\mu_1 \mu_2 - \lambda \mu_1 - \lambda \al (\mu_2 - \mu_1) }.
\end{equation}

From (\ref{eq:sum}), we also obtain the condition that the series is converging when $\lambda < \mu_2$, which is also the condition that the DTMC is an aperiodic irreducible Markov Chain, implying that the queue is stable.

However, since $\pi(0)$ denotes a probability, we additionally have the conditions $0 \leq \pi(0) \leq 1$. In order to fully characterize the previous condition in terms of $\lambda$, $\mu_1$ and $\mu_2$, we need to solve a polynomial equation of degree $M$, which will be evaluated numerically.

The average throughput of the secondary transmitter also depends on $\mathrm{Pr}\left(1 \leq Q \leq M \right) = \sum_{i=1}^M \pi(i)$, where
\begin{equation}
\mathrm{Pr}\left(1 \leq Q \leq M \right) = \frac{\lambda \left(1-\al \right)(\mu_2 - \lambda)}{\mu_1 \mu_2 - \lambda \mu_1 - \lambda \al (\mu_2 - \mu_1)}.
\end{equation}

Thus, the average throughout of the secondary node is given by
\begin{equation} \label{eq:T_s2}
T_{s} = \frac{(\mu_2 - \lambda) \left[(\mu_1 - \lambda)p_{2/2} + \lambda \left(1 - \al \right)qp_{2/1,2} \right]}{\mu_1 \mu_2 - \lambda \mu_1 - \lambda \al (\mu_2 - \mu_1)}.
\end{equation}

When the queue at the primary is stable then the aggregate throughput of the network in study is
\begin{equation} \label{eq:Taggr}
T_{aggr} = \lambda + T_s.
\end{equation}
In this work, we only consider one secondary transmitter-receiver pair, and the effect of the number of the secondary transmitters on the network performance will be investigated in a longer version of this paper.

\section{Numerical Results}\label{sec:Results}
In this section, we present numerical results for the throughput of the secondary node given in (\ref{eq:T_s2}) and the aggregate throughput of the network (\ref{eq:Taggr}) for different values of $M$, $q$ and $\lambda$. We consider the cases that the success probabilities between the transmitters and the receivers can be either high or low, and we consider strong and weak MPR capabilities for the receivers.

\begin{figure}[ht]
%\centering
\subfigure[Aggregate throughput vs. $q$]{
\includegraphics[scale=0.66]{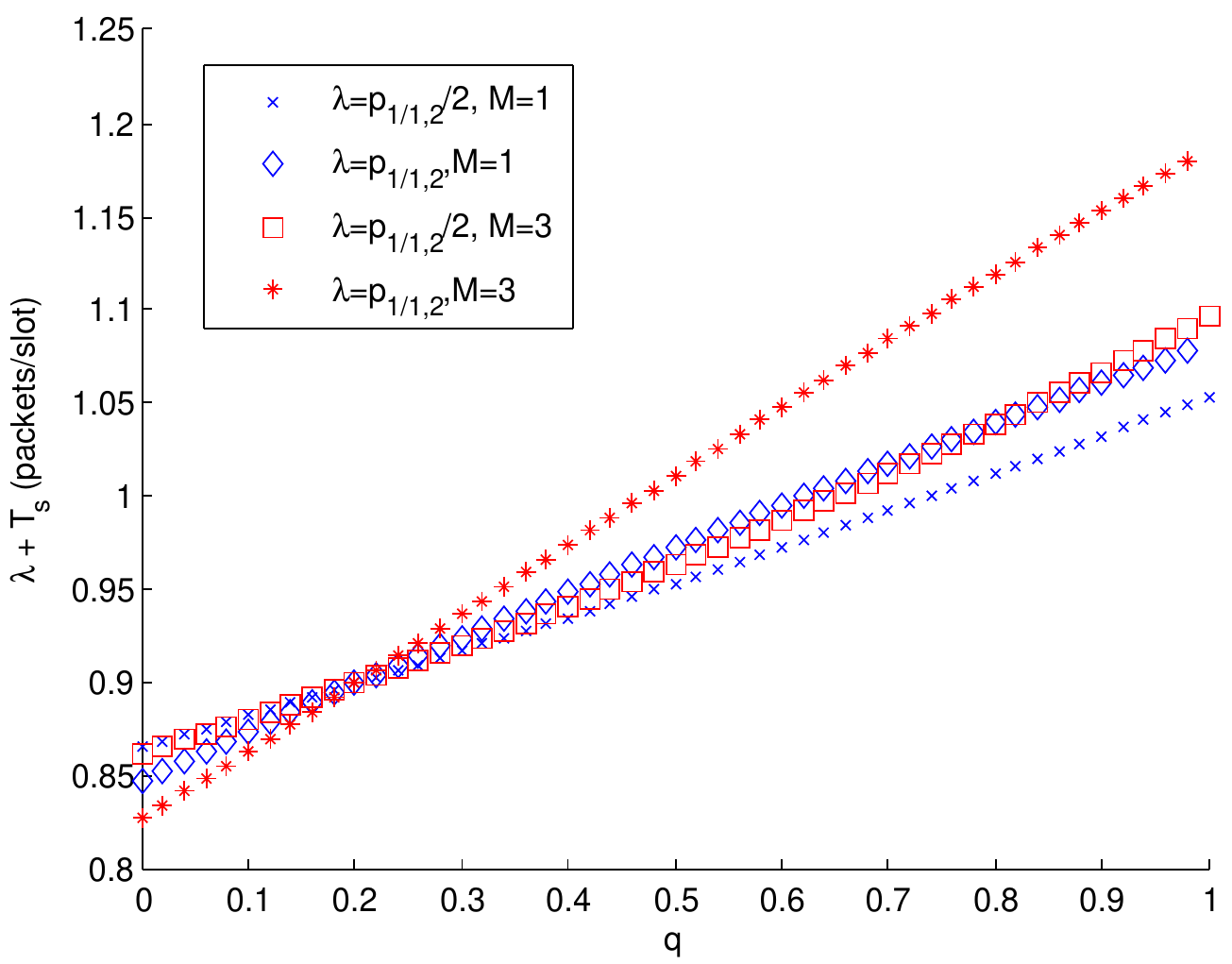}
\label{fig:thrvsq_good_high}
}
\subfigure[Aggregate throughput and throughput for the secondary transmitter vs. $\lambda$, $q=0.9$]{
\includegraphics[scale=0.66]{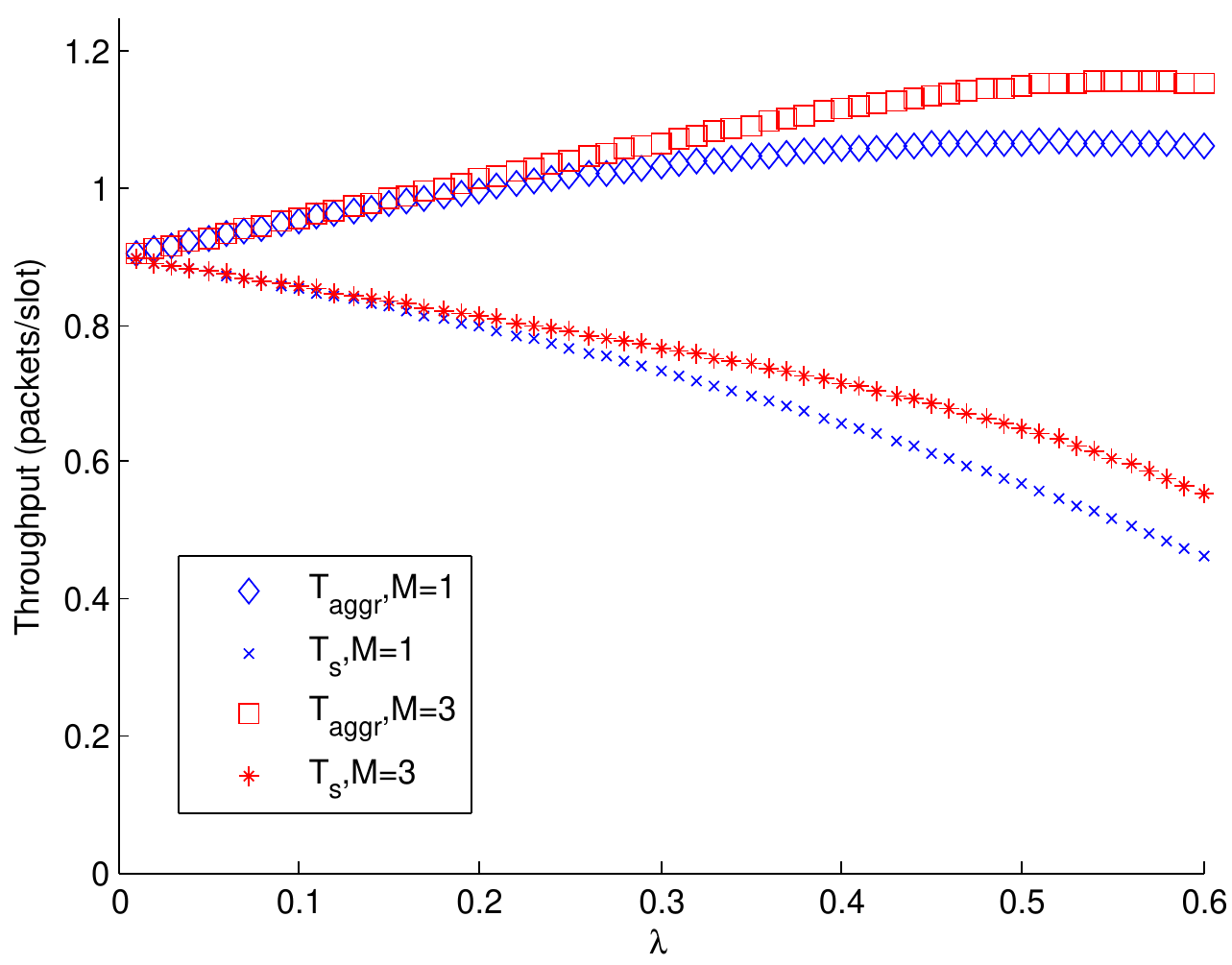}
\label{fig:thrvslambda_good_high}
}
\caption{High link success probabilities and strong MPR capabilities for the receivers: $p_{1/1}=0.8$, $p_{1/1,2}=0.6$, $p_{2/2}=0.9$ and $p_{2/1,2}=0.7$.}
\label{fig:thr_good_high}
\end{figure}

\begin{figure}[ht]
%\centering
\subfigure[Aggregate throughput vs. $q$]{
\includegraphics[scale=0.66]{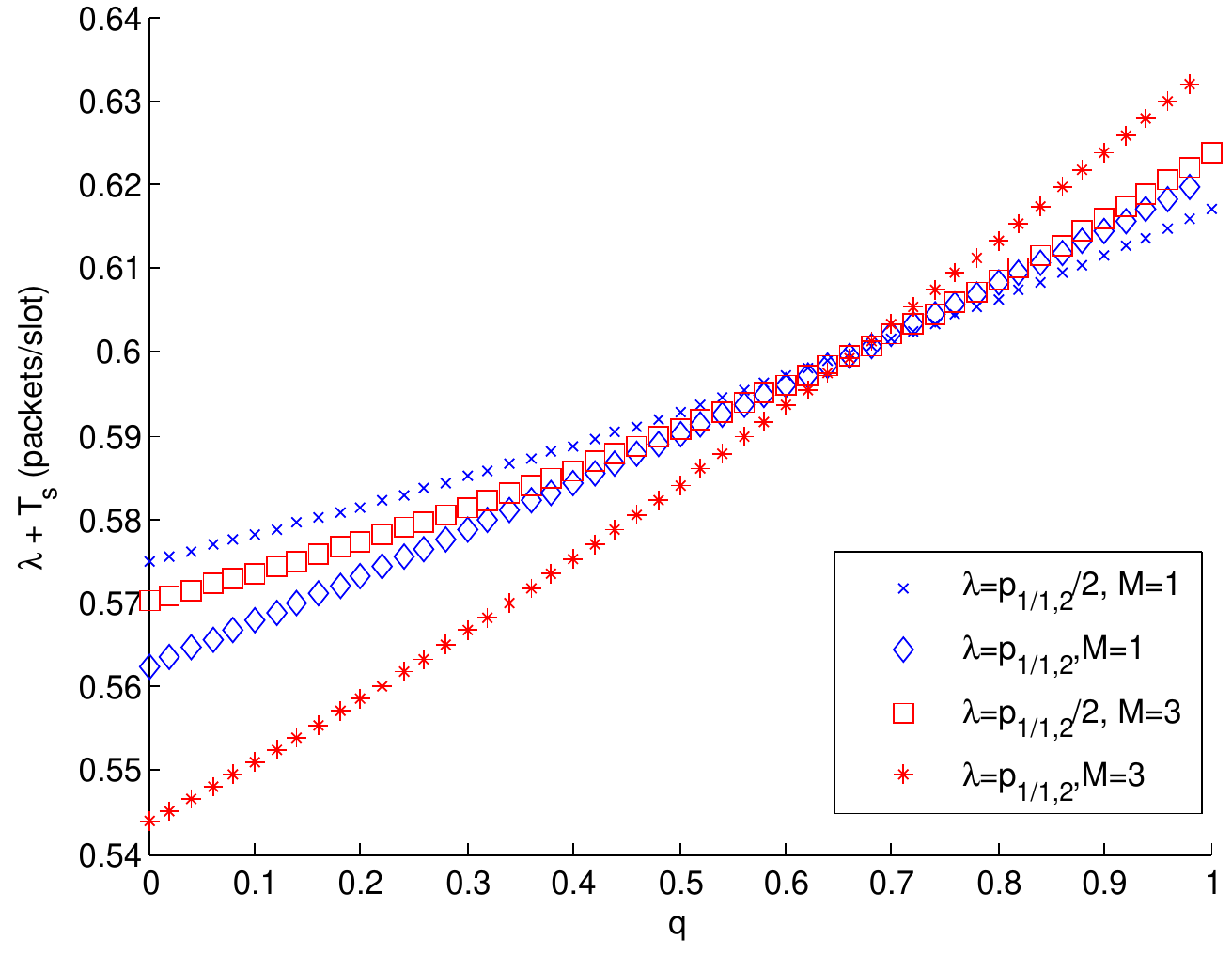}
\label{fig:thrvsq_bad_high}
}
\subfigure[Aggregate throughput and throughput for the secondary transmitter vs. $\lambda$, $q=0.9$]{
\includegraphics[scale=0.66]{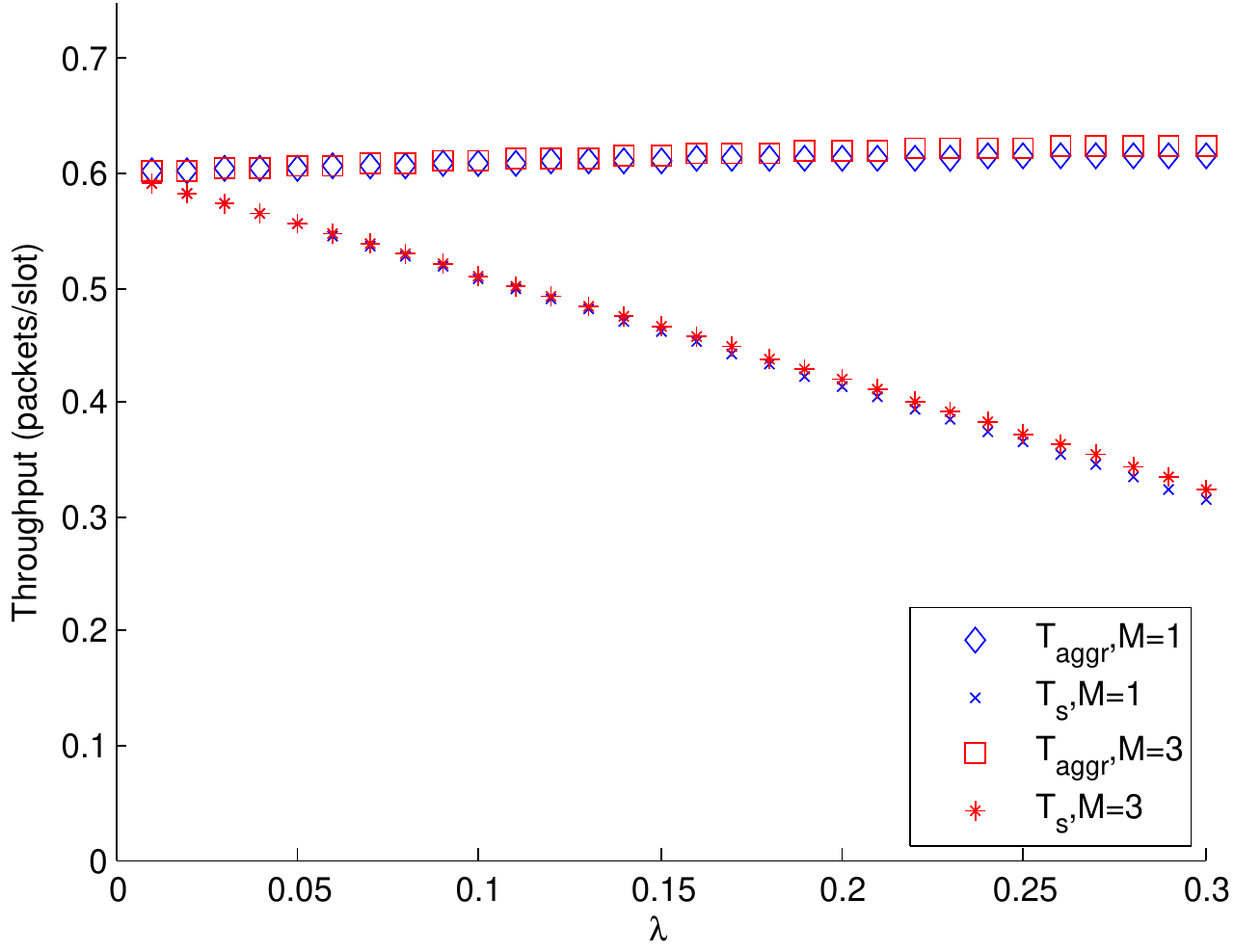}
\label{fig:thrvslambda_bad_high}
}
\caption{Low link success probabilities and strong MPR capabilities for the receivers: $p_{1/1}=0.5$, $p_{1/1,2}=0.3$, $p_{2/2}=0.6$ and $p_{2/1,2}=0.35$.}
\label{fig:thr_bad_high}
\end{figure}

In Figs.~\ref{fig:thr_good_high}-\ref{fig:thr_bad_high}, we illustrate the aggregate and the secondary node throughput versus the transmission probability $q$ and the arrival rate $\lambda$ for various values of congestion limit $M$ when the receivers have strong MPR capabilities. In this case, the throughput increases for $q$ and $\lambda$ increasing, otherwise it decreases. When $\lambda$ is relatively low, then $M$ does not really affect the system, since the probability that the queue is empty increases. Due to the low utilization in this case, choosing $M = 1$ in our protocol is beneficial. 
As illustrated in Fig.~\ref{fig:thrvslambda_good_high} and Fig.~\ref{fig:thrvslambda_bad_high}, the throughput of the secondary user decreases as $\lambda$ increases, whilst the aggregate throughput increases. This means that the decrease of secondary node throughput, due to the congestion of the primary node queue as we approach the saturation of the queue, is less than the increase of the $\lambda$.
Furthermore, when the channels between the transmitters and receivers have low success probabilities, then $M$ does not really affect the throughput.

\begin{figure}[ht]
%\centering
\subfigure[Aggregate throughput vs. $q$]{
\includegraphics[scale=0.6]{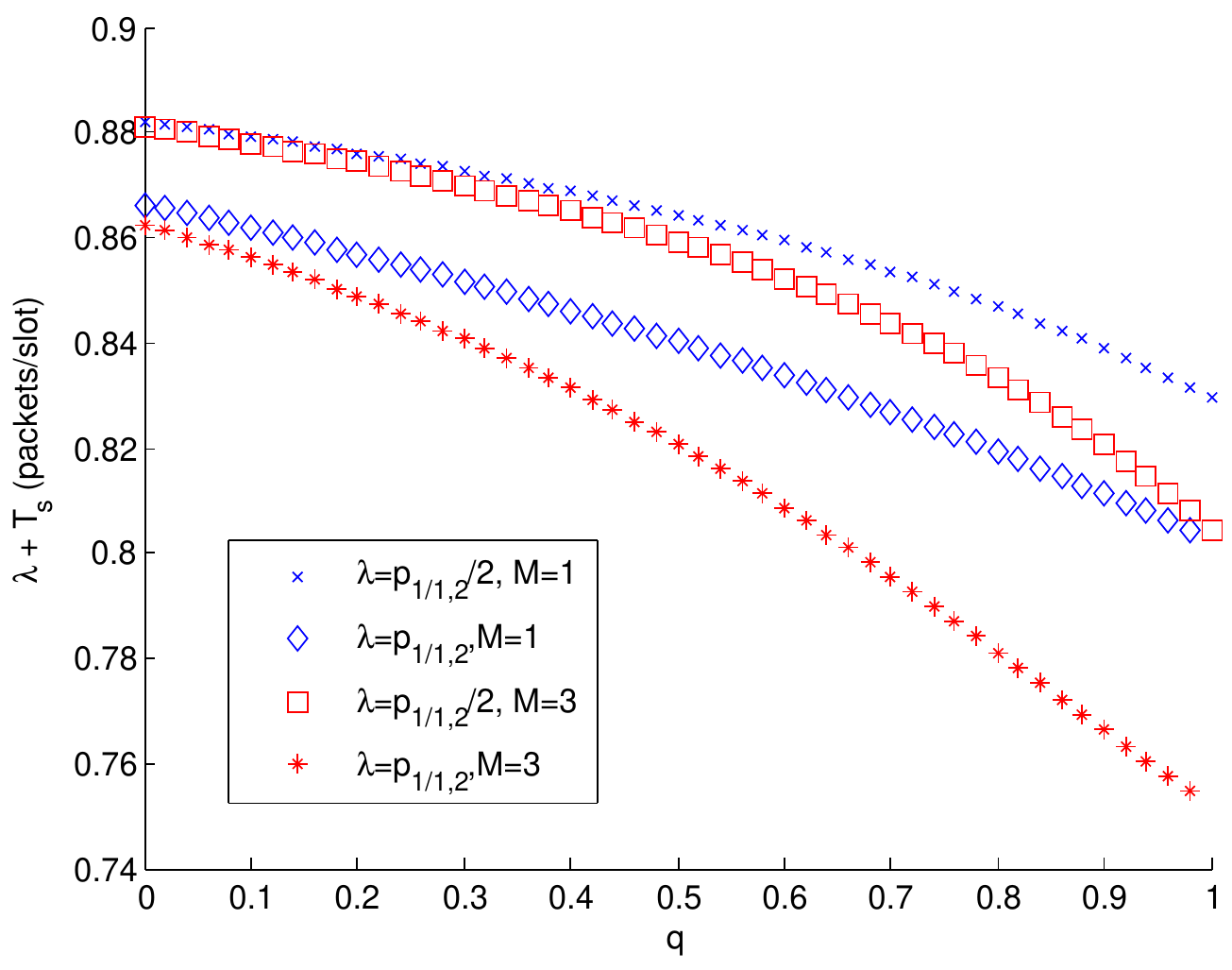}
\label{fig:thrvsq_good_low}
}
\subfigure[Aggregate throughput and throughput for the secondary transmitter vs. $\lambda$, $q=0.9$]{
\includegraphics[scale=0.6]{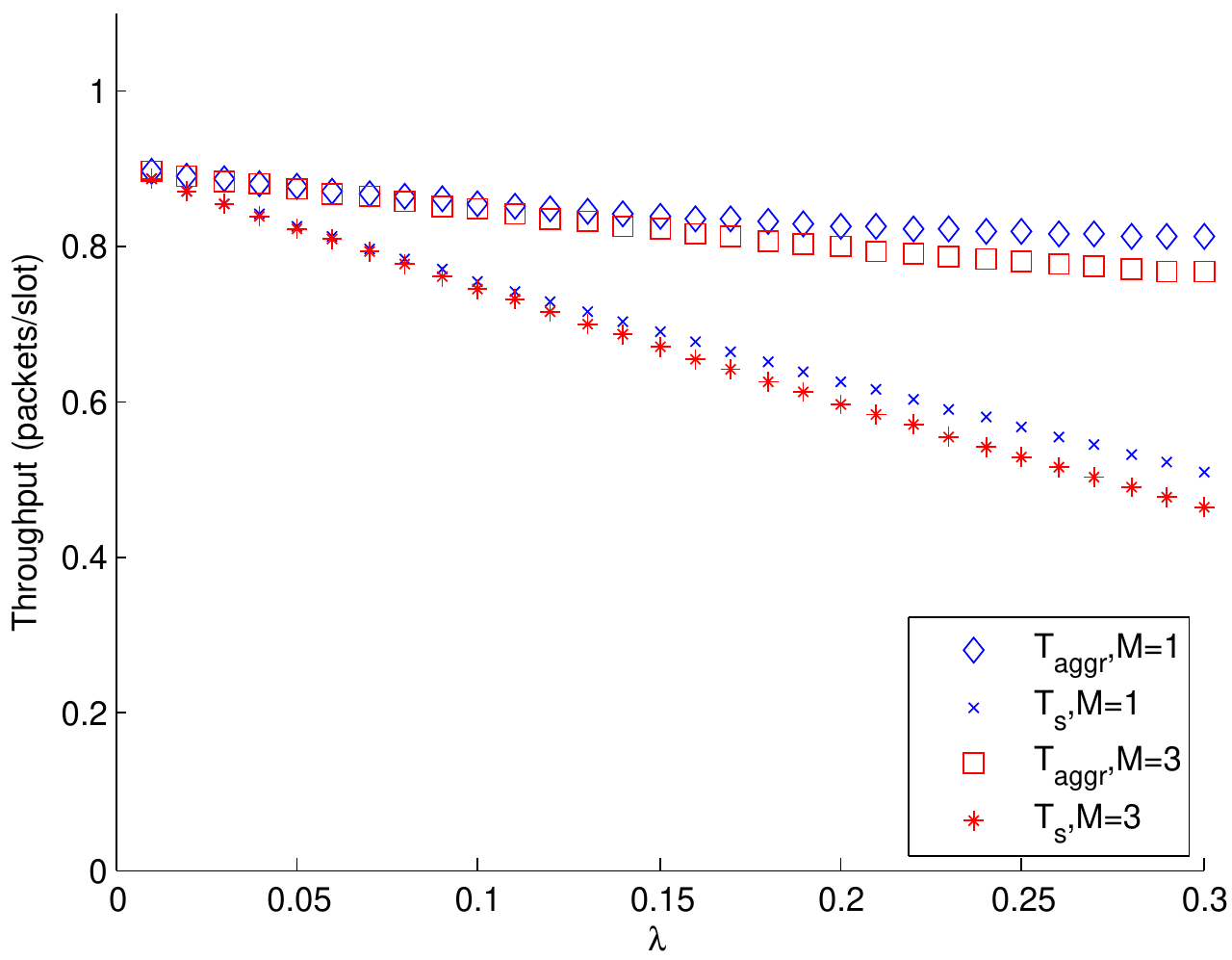}
\label{fig:thrvslambda_good_low}
}
\caption{High link success probabilities and weak MPR capabilities for the receivers: $p_{1/1}=0.8$, $p_{1/1,2}=0.3$, $p_{2/2}=0.9$ and $p_{2/1,2}=0.4$.}
\label{fig:thr_good_low}
\end{figure}

\begin{figure}[ht]
%\centering
\subfigure[Aggregate throughput vs. $q$]{
\includegraphics[scale=0.6]{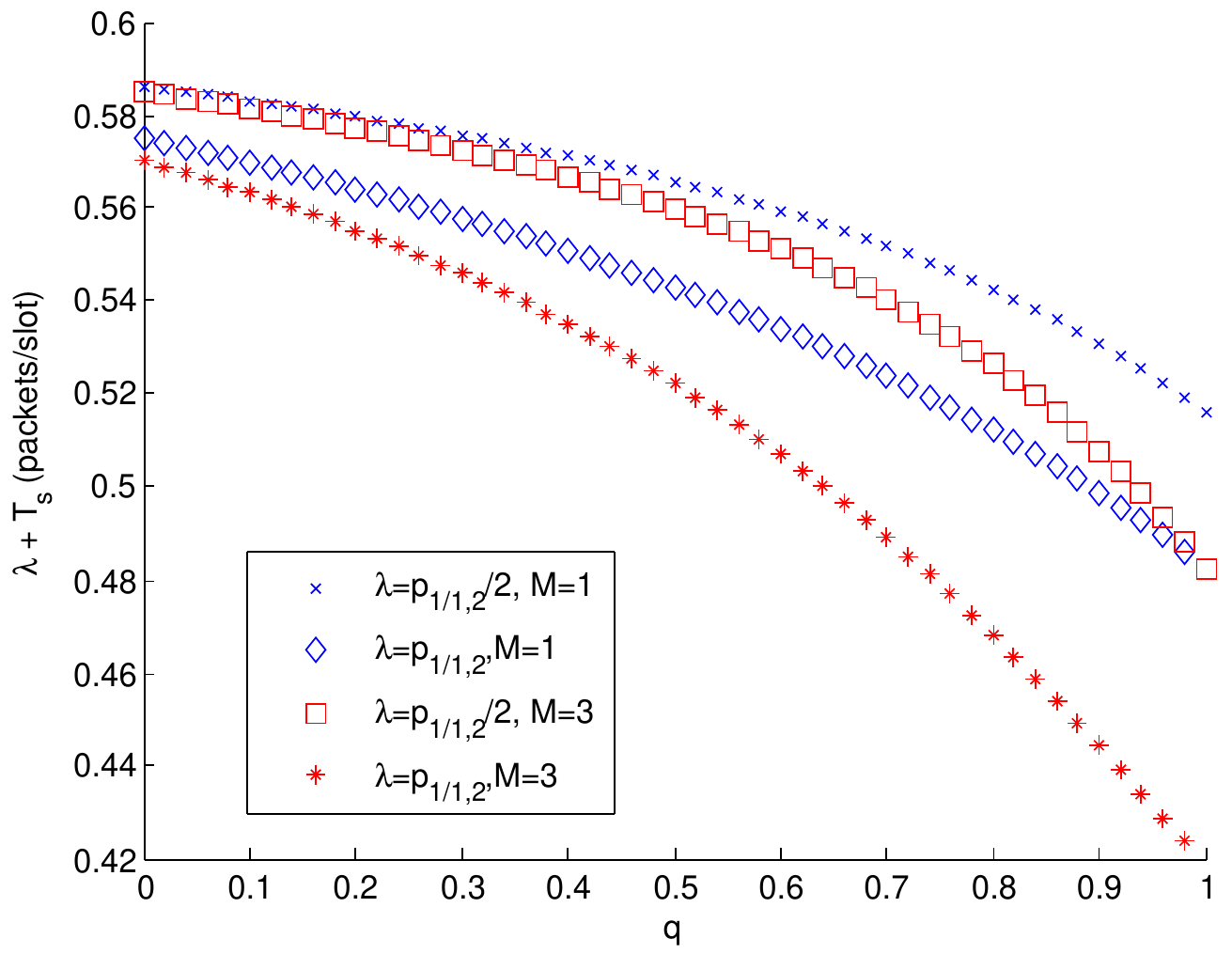}
\label{fig:thrvsq_bad_low}
}
\subfigure[Aggregate throughput and throughput for the secondary transmitter vs. $\lambda$, $q=0.9$]{
\includegraphics[scale=0.6]{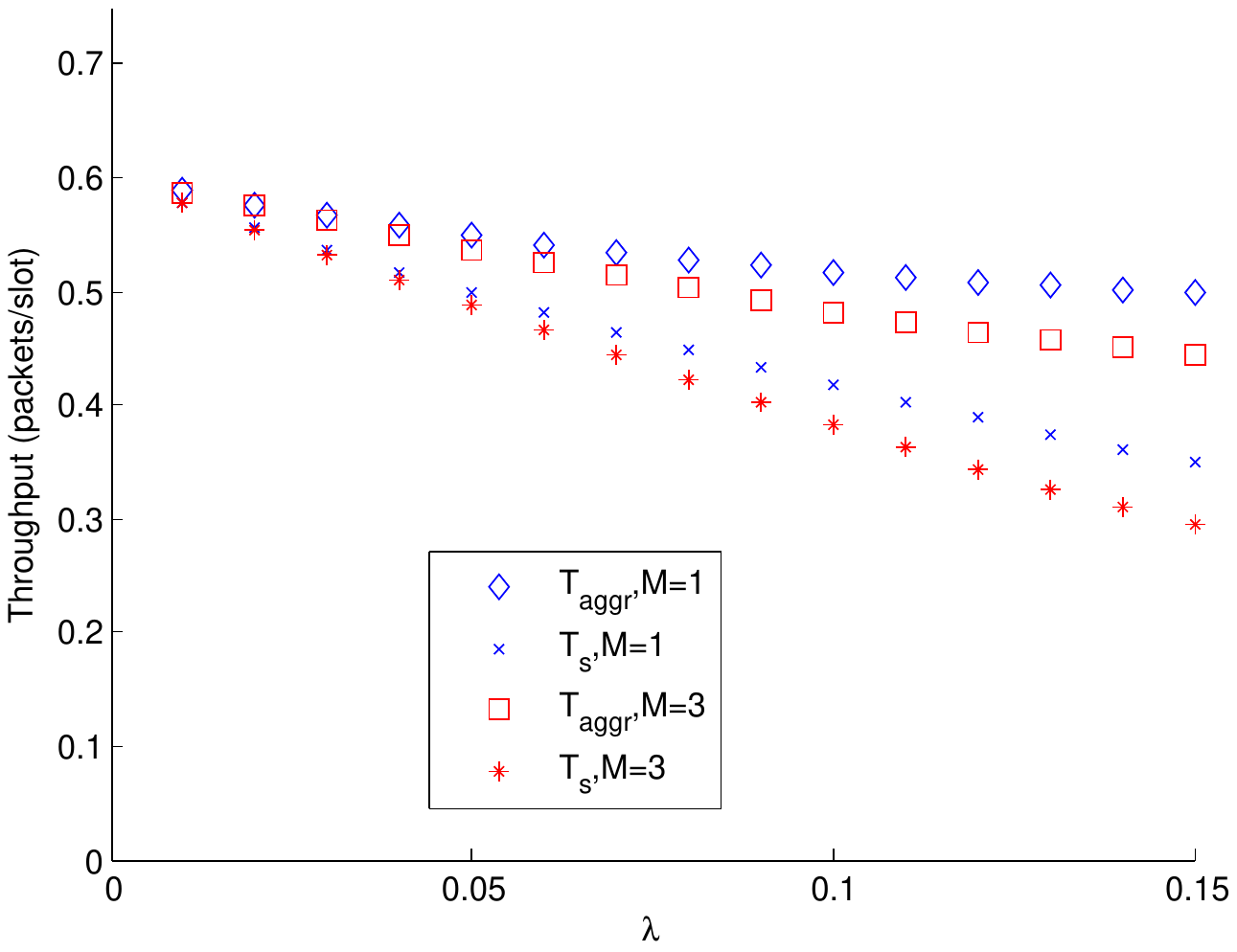}
\label{fig:thrvslambda_bad_low}
}
\caption{Low link success probabilities and weak MPR capabilities for the receivers: $p_{1/1}=0.5$, $p_{1/1,2}=0.15$, $p_{2/2}=0.6$ and $p_{2/1,2}=0.2$.}
\label{fig:thr_bad_low}
\end{figure}

In Figs.~\ref{fig:thr_good_low}-\ref{fig:thr_bad_low}, we depict the cases where the channels between the transmitters and the receivers are good and poor respectively, whereas the receivers have weak MPR capabilities. In that case, the throughput decreases as $q$ increases.
As $\lambda$ increases, both the secondary node and the aggregate throughput decreases. The drop of the secondary node throughput is bigger than the increase of the stable throughput of the primary node.

The main takeaway of our results is that if the receivers have strong MPR capabilities, the throughput increases as the transmission probability of the secondary increases, otherwise it decreases.

\section{Conclusions} \label{sec:conclusions}
In this work, we studied the performance of a cognitive network consisting of a primary and a secondary transmitter. The primary node has bursty arrivals, which are stored in its queue for a future transmission, while the secondary is assumed to be saturated. The secondary node transmits in a cognitive manner taking advantage of the emptiness of the primary node queue and in a random access when the queue is below a congestion limit $M$. We analyzed the performance of the primary node queue, and we obtained the stationary distribution and the stability conditions. We also derived the secondary node throughput, as well as the aggregate throughput as a function of the transmission probability of the secondary node, the arrival rate at the primary node, and the congestion limit $M$.

Further extensions of this work will include the delay analysis for this network and the dynamic adjustment of $M$ and $q$ depending on the arrival rate. Furthermore, the effect of the number of the secondary transmitters on the performance will be investigated. 

\bibliographystyle{IEEEtran}
\bibliography{thesis2}

\end{document}